\documentclass[11pt,preprint]{aastex}
\makeatletter
\def\lnyoro{\mathrel{\mathpalette\gl@align<}}
\def\gnyoro{\mathrel{\mathpalette\gl@align>}}
\def\gl@align#1#2{\lower.6ex\vbox{\baselineskip\z@skip\lineskip\z@\ialign{$\m@th
#1\hfil##\hfil$\crcr#2\crcr\sim\crcr}}}
\makeatother

\begin{document}

\title{\bf COLOR GRADIENTS IN EARLY-TYPE GALAXIES IN ABELL 2199}

\author{Naoyuki Tamura\altaffilmark{1}}
\affil{Department of Physics, University of Durham, Durham, DH1 3LE, UK}
\email{E-mail:naoyuki.tamura@durham.ac.uk}
\and
\author{Kouji Ohta\altaffilmark{1}}
\affil{Department of Astronomy, Kyoto University, Kyoto 606-8502, Japan}

\altaffiltext{1}{Visiting astronomer of the University of Hawaii
88$^{\prime\prime}$ telescope.}

\begin{abstract}

We performed $B$ and $R$ band surface photometry for E/S0 galaxies in a
nearby rich cluster ABELL 2199 to investigate their $B-R$ color
gradients ($d (B-R) / d \log r$). Our aims are to study statistical
properties of the color gradients and, by comparing them with those in
less dense environments, to examine environmental dependence of color
gradients in elliptical galaxies.
We studied the distribution of the $B-R$ color gradients in the cluster
ellipticals and found that the mean value of the color gradients is
$-0.09 \pm 0.04$ mag/dex, which can be converted to a metallicity
gradient ($d \log Z / d \log r$) of $\sim -0.3 \pm 0.1$. The gradient
seems to be comparable to that expected by a recent monolithic collapse
model.
We further studied the relations between the $B-R$ color gradients and
global properties of the galaxies. Our data suggest that for the
galaxies brighter than an $R$ band magnitude of $\sim 15$ mag, which is
roughly equivalent to $L^{*}$ at the distance of the cluster, brighter
galaxies tend to have steeper color gradients. Also, for the galaxies
with effective radii larger than $\sim 3^{\prime\prime}$, which nearly
corresponds to $L^{*}$ considering the correlation between galaxy
luminosity and effective radius for elliptical galaxies, the galaxies
with larger effective radii seem to have steeper color gradients. These
trends could appear if elliptical galaxies formed through the monolithic
collapse scenario.
On the contrary, it is found based on the published data that such
trends are not clearly seen for elliptical galaxies in less dense
environments, suggesting that elliptical galaxies in sparse environments
formed by galaxy mergers, though the distribution of the color gradients
is quite similar to that in the rich cluster.
In other words, our data and those in the literature suggest that there
is an environmental dependence in the relationship between color
gradient and global properties of elliptical galaxies, while the
distribution of the values of color gradients is nearly independent of
galaxy environment.
These results do not fully support the view that ellipticals in rich
clusters formed through the monolithic collapse while those in sparse
environments formed through galaxy mergers, because the latter
ellipticals are expected to have color gradients shallower on average
than the former.
This apparent conflict may be reconciled by taking into account star
formation and subsequent chemical enrichment induced by galaxy merger,
which may allow a merger remnant to acquire a metallicity gradient and
thus compensate the dilution of the existing metallicity gradients in
the progenitors by mixture of stars during the galaxy merger.

\end{abstract}


\section{INTRODUCTION}

Elliptical galaxies which reside in a cluster show a tight
color-magnitude (CM) relation; brighter ellipticals are systematically
redder than fainter ones (e.g., Bower, Lucey, \& Ellis 1992). Recently,
it has been found that the relation still holds even in distant rich
clusters up to $z \sim 1$ (e.g., Stanford, Eisenhardt, \& Dickinson
1998). Based on these observational results, galaxy evolution models
taking into account chemical evolution claim that ellipticals in rich
clusters formed at high redshift (e.g., $z > 3$) by a monolithic
collapse of a gas cloud (e.g., Kodama \& Arimoto 1997; Kodama et al.
1998). However there is another formation scenario for elliptical
galaxies, i.e., the hierarchical merging scenario, which may also explain
these observational results (e.g., Kauffmann \& Charlot 1998; Shioya \&
Bekki 1998). Which of the scenarios dominates the formation of
elliptical galaxies has been one of the controversial issues. In order
to address this problem, observational properties representing histories
of dynamical evolution such as merging histories as well as histories of
chemical evolution of galaxies should be focused on. One of the
promising candidates is a radial variation of stellar metallicity, i.e.,
a metallicity gradient, in an elliptical galaxy.

Metallicity gradients in elliptical galaxies are expected to represent
how the galaxies have assembled. {\it In the monolithic collapse
scenario}, a metallicity gradient in an elliptical galaxy can be built
up by a dissipative collapse in its formation phase associated with a
burst of star formation and a subsequent galactic wind, because of a
more extended period of active star formation and more metal enrichment
in the inner region (Larson 1974a, b; Carlberg 1984; Martinelli et
al. 1998; Tantalo et al. 1998; Kawata 1999). In addition, more massive
ellipticals are predicted to have steeper gradients (Larson 1974b;
Carlberg 1984).
{\it In the hierarchical merging scenario}, stellar populations in the
progenitor galaxies are expected to be well mixed in major mergers and
the existing metallicity gradients are likely to be diluted (White
1980). Although metallicity gradients may be newly produced due to
secondary star formations induced by mergers (e.g., Mihos \& Hernquist
1994), these gradients are suggested to be appreciably shallower than
those expected by the monolithic collapse scenario (Bekki \& Shioya
1999). It should be mentioned that in this scenario, the metallicity
gradients would be different from galaxy to galaxy due to their
different histories of mass assembly and star formation. Therefore, no
strong correlation between metallicity gradient and galaxy mass would be
expected. Although more massive ellipticals might have shallower
gradients because they are expected to experience more galaxy mergers in
the hierarchical merging scheme and their metallicity gradients may be
smeared out more seriously, it is opposite to the correlation expected
by the monolithic collapse scenario.

Evolutionary histories of elliptical galaxies can be dependent on galaxy
environment. The presence of tight CM relations in rich clusters at high
redshift seems to support the monolithic formation at, e.g., $z \geq 3$
for ellipticals in such high density environments (e.g., Kodama et
al. 1998). Even in the Cold Dark Matter (CDM) universe, high peaks of
density fluctuations at such high redshifts could form giant galaxies
and be related to clusters of galaxies along the biased galaxy formation
scenario (Peacock 1999 and references therein). Since it is expected
that the mean density of matter is high and objects are gas-rich at
those epochs, formations of galaxies are predicted to be
accelerated. This implies that elliptical galaxies in dense environments
formed through bursts of star formation at high redshift, which mimic
the initial starbursts expected in the monolithic collapse scenario, and
tend to reside in rich clusters.
On the other hand, the field corresponds to the environment without such
high peaks of density fluctuations. The density fluctuations are
expected to evolve more slowly and collapse at lower redshifts. Galaxies
would evolve more quiescently through bottom-up processes of smaller
sub-galaxies and could settle down to disk galaxies preferentially. In
this case, elliptical galaxies are more likely to form through major
mergers of disk galaxies at low redshifts (e.g., $z < 2$).
Indeed, some recent observational or theoretical studies suggest that
elliptical galaxies which formed through major mergers tend to inhabit
in field environments and they have experienced mergers inducing star
formation until recent epochs such as $z \lnyoro 1$ (e.g., Baugh et al.
1996; Governato et al. 1999; Kauffmann 1996; Menanteau et al. 1999;
Menanteau, Abraham, \& Ellis 2001). Such late star forming activities
could make their luminosity-weighted ages at $z = 0$ systematically
younger (Terlevich \& Forbes 2002).

In summary, if we consider that elliptical galaxies in cluster
environments are suggested to have formed through a process like the
monolithic collapse while those in sparse environments are through major
mergers, the formers are predicted to have steeper metallicity gradients
on average than the latter. Also, a clear correlation between
metallicity gradients and global properties of galaxies (mass, size,
luminosity, etc) is not expected for ellipticals in sparse environments
but expected for cluster ellipticals. To see whether observational data
show these kinds of environmental dependence or not will provide strong
constraints on the formation processes of galaxies and the roles played
by galaxy environments in evolutionary histories of galaxies.

Although the above issues motivate us to examine metallicity gradients
in elliptical galaxies, the problem is how the gradients are to be
examined observationally. It is known that nearby elliptical galaxies
have color gradients; colors in an elliptical galaxy gradually become
bluer with increasing radius (e.g., Vader et al. 1988; Franx,
Illingworth, \& Heckman 1989; Peletier et al. 1990a; Peletier,
Valentijn, \& Jameson 1990b; Goudfrooij et al. 1994; Michard 1999;
Idiart, Michard, \& de Freitas Pacheco 2002). Although the
age--metallicity degeneracy prevents us from interpreting the origin of
the color gradients to be the metallicity gradients, an evolution of a
color gradient with redshift has been studied to disentangle the
degeneracy and it is concluded that the primary origin of color
gradients in elliptical are not the age gradients but the metallicity
gradients both for the elliptical galaxies in field environments and for
those in cluster environments (Tamura et al. 2000; Tamura \& Ohta 2000;
Saglia et al. 2000; see also Hinkley \& Im 2001).
It should be mentioned here that dust extinction may have some effects
on a color gradient in an elliptical galaxy (Goudfrooij \& de Jong
1995). As a matter of fact, if an elliptical galaxy consists of a
mixture of stars without any population gradients and diffusely
distributed dust, a calculation of the radiative transfer within the
galaxy suggests that the color gradients could be reproduced only by the
dust effects (Witt, Thronson, \& Capuano 1992; Wise \& Silva 1996).
However, many elliptical galaxies show radial gradients of metal
absorption line index such as Mg$_{2}$, Fe$_{1}$(5270 \AA) and
Fe$_{2}$(5335 \AA) as well as the color gradients (e.g., Carollo,
Danziger, \& Buson 1993; Davies, Sadler, \& Peletier 1993; Gonz\'{a}lez
1993; Kobayashi \& Arimoto 1999; Mehlert et al. 2000), and the radial
gradients of the metal absorption line indices are unlikely to be
created by dust extinction.
It has also been suggested that {\it on average}, metallicity gradients
in elliptical galaxies as estimated by a population synthesis model from
color gradients are consistent with those estimated from absorption line
index gradients (e.g., Peletier et al. 1990a; Davies et al. 1993),
although there are exceptions. We thus assume that metallicity effects
seem to be the prime cause of color gradients in elliptical galaxies.
Although both color gradients and metal absorption line strength
gradients can be used to trace the metallicity gradients, we study color
gradients in this paper because broad band imaging observations are much
easier to obtain high $S/N$ data to study the gradients in elliptical
galaxies and to enlarge samples including further extension towards high
$z$ in future.

The previous studies suggest that the color gradients in nearby
elliptical galaxies do not correlate with any global properties (mass,
luminosity, etc) of elliptical galaxies (e.g., Peletier et al. 1990a).
Kobayashi \& Arimoto (1999) presented the same results by compiling the
data of the Mg$_{2}$ gradients in the literature and constructing a
large sample of nearby elliptical galaxies. These studies may support
the merging scenario.  However, it should be stressed that most of the
ellipticals studied reside in low density environments such as field or
group. Although some elliptical galaxies in Virgo cluster have been
studied (Peletier et al. 1990a; Goudfrooij et al. 1994; Michard 1999;
Idiart et al. 2002), Virgo cluster is not a rich cluster but a poor
cluster with a rather sparse spatial distribution of galaxies.
Therefore, it is unlikely to be sufficient to know whether there is a
dependence of color gradient on galaxy environment or not, and one of
the straightforward extensions to address this issue is to study
elliptical galaxies in rich clusters.

In this study, we examine color gradients in elliptical galaxies in a
nearby rich cluster, ABELL 2199. It is known that this cluster is
significantly richer than Virgo cluster and is a spiral poor and
centrally concentrated cD cluster (e.g., Butcher \& Oemler 1985). It is
also considered to be a relaxed cluster based on its regular and
symmetrical appearance in X-ray (e.g., Buote \& Canizares 1996).
Although Coma cluster is a good candidate as a nearby typical rich
cluster and global properties of the member galaxies have been well
studied, it has been pointed out to be a merging cluster with two X-ray
peaks (e.g., Caldwell et al. 1993; Burns et al. 1994) and thus the
galaxies in Coma cluster may have been also disturbed by recent galaxy
mergers. We consider that elliptical galaxies in a more regular and
relaxed cluster should be studied at first.

The layout of this paper is as follows. Observation and data reduction
are described in the next section. In \S~3, sample definition is
described and radial profiles of surface brightness and color of the
sample galaxies are presented. Also, color gradients in the sample
galaxies are derived and their statistical properties are
investigated. We discuss implications of the results in \S~4, and
summarize this paper in \S~5. We assume $H_0 = 50$ km s$^{-1}$
Mpc$^{-1}$ throughout this paper.

\section{OBSERVATION AND DATA REDUCTION}

Imaging observations in $B$ and $R$ bands were made on June 20 and 21,
2001, with Tektronix 2K$\times$2K CCD on the University of Hawaii 2.2m
telescope. One pixel of the CCD chip corresponds to
$0_{\cdot}^{\prime\prime}22$ and thus each frame covers an area of about
$7_{\cdot}^{\prime}5\times7_{\cdot}^{\prime}5$. Observation log is
presented in Table \ref{list1}. Although early-type (E, E/S0, and S0)
galaxies in ABELL 2199 were primary targets in this observing run, the
central region of ABELL 2634 was also observed as a pilot study to
extend our sample to other clusters. 
(ABELL 2634 is also a centrally concentrated cluster but slightly less
rich than ABELL 2199 (e.g., Pinkney et al. 1993).)
In addition, 3 elliptical galaxies (NGC 5638, NGC 5831, and NGC 7619),
of which color gradients were examined in the previous studies, were
observed for comparison. A typical seeing size during the observing run
was $\sim 0_{\cdot}^{\prime\prime}9$ in FWHM. Total exposure time at
each field is typically 1,800 sec for $B$ band and 750 sec for $R$ band,
each of which was divided into three or four exposures with dithering.

All of the imaging data were reduced with IRAF\footnote{IRAF is
distributed by the National Optical Astronomy Observatories, which is
operated by the Association of Universities for Research in Astronomy,
Inc. under cooperative agreement with the National Science Foundation}
in the standard procedure. After bias subtraction, flat-fielding by
dome-flat frames, and sky subtraction were carried out in each frame,
both of the $B$ and $R$ band images at each field were registered by
sub-pixel shifts. Next, slight differences in flux among the images in
each band were corrected by multiplying an image by a constant, and the
FWHMs of the point spread functions (PSFs) were matched by convolutions
of images with Gaussian kernels. These images were combined with the 3
$\sigma$ clipping algorithm. After a stacked image was made in each
band, a PSF size in one band was matched to that in the other band by
smoothing with Gaussian kernels. A resulting PSF size is typically
$1_{\cdot}^{\prime\prime}1$ in FWHM.

The photometric calibration was done by using the standard stars in
Landolt (1992). Aperture photometry was carried out after a subtraction
of a local residual sky background around each galaxy, which was
determined by ``mode'' in an annulus with an inner radius of $\sim
3R_{25}$ and a width of $11^{\prime\prime}$ (50 pix).
In order to check the accuracy of the photometry and sky subtraction,
growth curves of the objects were compared with those taken from
HYPERCAT\footnote{http://www-obs.univ-lyon1.fr/hypercat}: several
galaxies in the clusters as well as NGC 5638, NGC 5831, and NGC 7619
have the data of aperture photometry. 
It is found that there seem to be zero-point offsets between our growth
curves and those taken from HYPERCAT, though the amounts of the offsets
are $0 - 0.1$ mag in each band. This may imply that the observational
condition was slightly non-photometric. However, if we correct for the
offsets in each field, a tight CM relation consistent with that observed
in Coma cluster is obtained both in ABELL 2199 and in ABELL 2634. We
calculated an average amount of the zero-point shift on each night and
it is corrected for by a constant offset in the following analyses.
Since we focus on a color gradient in a galaxy in this study, such
zero-point offset does not cause any serious effects.

\section{DATA ANALYSES AND RESULTS}

\subsection{Sample Definition}

The target fields were chosen based on the catalogs of early-type
galaxies in the clusters by Lucey et al. (1997). 40 early-type (E, E/S0,
and S0) galaxies in ABELL 2199 and 11 early-type galaxies in ABELL 2634
are included in the fields; most of the galaxy morphology is originally
from Butcher \& Oemler (1985) and Rood \& Sastry (1972), while for
several galaxies which had not been classified in the literature, we
provided morphological types to the galaxies based on eye inspection.
Although we may be able to study color gradients of these
morphologically selected galaxies, it is worth noting that there is a
variety in spheroidal-to-total luminosity ratio among galaxies in each
morphological type (Simien \& de Vaucouleurs 1986). In addition, since
the clusters studied here are relatively distant, visual classifications
of morphology of fainter galaxies may be less reliable.
In order to reduce contamination of disk dominated galaxies in a sample,
we made azimuthally averaged radial surface brightness profiles in $B$
and $R$ bands of each galaxy, and performed decompositions of the
profiles into bulge and disk components to estimate a bulge-to-total
luminosity ratio ($B/T$).

The radial surface brightness profiles are obtained along ellipses
fitted to the isophotes with the ELLIPSE task in the STSDAS package with
a radial sampling of 0$_{\cdot}^{\prime\prime}$22 (1 pix) along major
axis. In fitting an ellipse to an isophote, ellipticity is set to be a
free parameter. Galaxy center is fixed to a centroid of a galaxy in the
$R$ band image and position angle is also fixed to that of an ellipse
within which half of the galaxy light is included. Even though position
angle is allowed to be a free parameter, no significant differences in
the surface brightness profiles appear. In the following analyses, all
of the radial profiles and the parameters related to radius are
expressed in terms of the equivalent radius of an ellipse: $\sqrt{ab}$,
where $a$ and $b$ are the semi-major and semi-minor axes of the ellipse.
The ELLIPSE task employs an iterative algorithm of rejecting pixels
significantly deviant from an average on each ellipse. This enables us
to eliminate effects of contamination of background/foreground objects
at least to some extent. However, since bright objects in the neighbor
of a target galaxy may have serious effects on the fitted ellipses, the
neighbors are masked out beforehand. It is found that effective radii
and color gradients are nearly invariant to whether these masking
processes are applied or not and the differences are comparable to or
smaller than errors in the individual measurements of the parameters.
In practice, some galaxies are ``embedded'' in an envelope of a large
galaxy in the neighbor. In order to dig a galaxy out of the envelope, we
model the neighboring galaxy using the BMODEL task in the STSDAS package
and subtract it. In most cases, an estimated effective radius and color
gradient in such an ``embedded'' galaxy only slightly depend on whether
the subtraction process is applied or not, though there are a few
galaxies in which the discrepancies can amount to $\sim 50$\% in
effective radius and $-0.05$ mag/dex in color gradient. Although we
include galaxies in the following analyses and adopt the parameters
estimated after the subtraction processes, they are flagged in the
results to show that the subtraction processes are employed and the
measurements of effective radii and color gradients may be more
uncertain.

The decomposition was performed by fitting a surface brightness profile
with the sum of an $r^{1/4}$ law bulge and an exponential disk profiles
so as to minimize $\chi^2$. We considered seeing effects on the surface
brightness profiles using the formulation by Pritchet \& Kline (1981).
We chose to sample early-type galaxies of which $B/T$s in $B$ band are
estimated to be larger than 0.6. This criterion isolates galaxies
earlier than E/S0 or $T \leq -3$ according to Simien \& de Vaucouleurs
(1986), where $B/T$s of nearby galaxies are also derived in $B$ band. In
this selection procedure, 31 galaxies are picked out in ABELL 2199,
while $\sim 13$\% (3/25) of the ellipticals (E and E/S0) and $\sim 40$\%
(6/15) of the S0 galaxies among the 40 ellipticals and S0s were
excluded. In ABELL 2634, 7 galaxies are picked out and two of the
elliptical galaxies and one of the S0 galaxies were excluded. These
fractions are consistent with those expected from the dispersion of the
relation between morphological type and $B/T$ suggested by Simien \& de
Vaucouleurs (1986). It is suggested according to the decomposition that
in almost all the sample galaxies, spheroidal components dominate in
surface brightness at radii where radial profiles of color are obtained
reliably and color gradients are calculated. However, for two of the
galaxies in ABELL 2199, their surface brightness profiles suggest that
disk light starts dominating at a radius smaller than the outer cut-off
radius in deriving a color gradient (see \S~\ref{clg} for detail of
outer cut-off radius).  Therefore, we excluded these two galaxies.
Hence, the sample in ABELL 2199 consists of 29 early-type galaxies.
The R band images of the sample galaxies are shown in the top panels of
Figure 1, and their observational properties are listed in Table
\ref{list2}. The total magnitudes were obtained by using SExtractor
(Bertin \& Arnouts 1996) and the ``BEST'' magnitudes were adopted.  A
$B-R$ color was calculated within an aperture whose radius is set to an
effective radius of the galaxy in question. A central velocity
dispersion ($\sigma_0$) and an Mg$_2$ index are taken from Lucey et
al. (1997).

The surface brightness profiles of these sample galaxies are shown in
the middle panels of Figure 1. An error bar attached to each of the data
points includes an photometric error, a local sky subtraction error, and
a dispersion of surface brightness along each fitted ellipse. A local
background around each object and its subtraction error were estimated
by calculating a standard deviation around the mode in an annulus with
an inner radius of $\sim 3R_{25}$ and a width of $11^{\prime\prime}$ (50
pix). In Figure 1(a) and 1(b), the sample galaxies in ABELL 2199 and
those in ABELL 2634, respectively, are presented in order of $R$ band
total magnitude. The three galaxies (NGC 5638, NGC 5831, and NGC 7619)
which were taken for calibration purposes are shown in Figure 1(c).
Effective radii ($r_{e}$) of the sample galaxies were derived by
regression lines best fitted to the $R$ band surface brightness
profiles.
In fitting a regression line, the data points in the outermost regions
of which $S/N$s are low were removed beforehand. In addition, the
innermost regions probably flattened by the seeing effects were not
used. The resulting fitting regions are shown by the two vertical solid
lines in the middle panels of Figure 1.  The effective radii obtained
are presented in Table \ref{list2}. It is mentioned that if the
innermost regions are removed properly, they are almost insensitive to
the inner and outer cutoff radii.
The effective radii correlate with the $R$ band total magnitudes: more
luminous galaxies have larger effective radii, which is related to the
Fundamental Plane. (This correlation is referred to ``scaling relation''
hereafter.) Most of the sample galaxies have $R$ band total magnitudes
between 14 mag and 16 mag, and in this magnitude range an effective
radius typically varies from 6$^{\prime\prime}$ to $1^{\prime\prime}$.

\subsection{Two Dimensional Color Distributions and Radial Color Profiles}

Before obtaining radial color profiles in the sample galaxies, it is
worth addressing any trends seen in their two dimensional color
distributions. We made the $B-R$ color maps of the galaxies after the
seeing sizes in both of the images were matched. In Figure 2(a) and
2(b), the color maps of the sample galaxies in ABELL 2199 and ABELL
2634, respectively, are presented in order of $R$ band total magnitude
of a galaxy from left to right and from top to bottom. The three
calibration galaxies (NGC 5638, NGC 5831, and NGC 7619) are presented in
Figure 2(c). A cross mark in each panel shows the centroid of a galaxy
in $R$ band. The regions with $B$ band surface brightness brighter than
$\sim$ 23 mag/arcsec$^2$ in the galaxies are indicated in these color
maps and Gaussian smoothing with a kernel of $\sim
0_{\cdot}^{\prime\prime}3$ was performed for the clearer
presentation. These are the {\it pseudo} color maps; a range of color
adopted in each map is shown by the wedge at the right side of the panel
and is different from galaxy to galaxy.  We note that the features like
vertical lines seen in BO 24 and NGC 7720 are caused by bad columns on
the CCD surviving due to a small dithering scale in the horizontal
direction at the fields. It is found that almost all the galaxies show
axisymmetric color distributions independent of morphology and
luminosity. This demonstrates that their radial color profiles and the
color gradients represent the color distributions well.

The radial profiles of $B-R$ color of the sample galaxies are shown in
the bottom panels in Figure 1. Each color profile was constructed by
subtracting the surface brightness profile in an $R$ band image from
that in a $B$ band image. Both of these surface brightness profiles were
made along the same ellipses fitted to isophotes in the $R$ band image.
An error bar attached to each data point in the color profile includes a
photometric error, a local sky subtraction error, and a dispersion of
colors along each elliptical isophote.
Most of these color profiles between the inner and outer cut-off radii
(see next subsection for detail) seem to be linear, while some of them
appear to bend in the outermost regions. Deeper data are essential to
confirm whether the bends are real or not and to define a color profile
of a galaxy out to a larger radius.

\subsection{Color Gradients} \label{clg}

A slope of a radial color profile, that is, a color gradient in each
galaxy is derived by the least square fitting method using a portion of
a color profile between an inner and outer cut-off radii, which are
defined as follows.

The outer cut-off radius is defined to be a radius at which $R$ band
surface brightness of a galaxy is 22.5 mag/arcsec$^2$. This typically
corresponds to a radius at which an error of $B-R$ color amounts to
$\sim$ 0.1 mag. An alternative definition of the outer cut-off radius
could be effective radius. However, since a less luminous early-type
galaxy tends to have a smaller effective radius, the number of data
points available to fit a regression line to derive a color gradient is
significantly reduced and thus the number of galaxies in which color
gradients are derived reliably is also reduced. If a threshold in
surface brightness is adopted to define the cut-off radius, the effects
of the reductions are mitigated. It should be stressed here that the
color profiles in most of the sample galaxies are well approximated by
lines even if the regions beyond their effective radii are included. For
these reasons, we have chosen to use not an effective radius but a
radius defined by the surface brightness for the outer cut-off radius.

The inner cut-off radius is necessary to avoid seeing effects on color
profiles. Colors in the inner regions of a galaxy are affected more
seriously by seeing than those in the outer regions, which can cause
some amount of change in color gradient. In order to understand seeing
effects on color gradients in our data and to define the inner cut-off
radius optimal for our study, galaxies simulated with the MKOBJECT task
in the ARTDATA package are put onto blank fields in the stacked
images. The model galaxies are assumed to have de Vaucouleurs law
profiles, each of which is convolved with a Gaussian kernel with the
same FWHM as that in the image. An integrated $B-R$ color in a model
galaxy is assumed to be 1.5 mag. An ``intrinsic'' color gradient is
given to a model galaxy by adopting a smaller effective radius in $R$
band than that in $B$ band, and three cases of color gradient ($d (B-R)
/ d \log r = -0.1, -0.2$, and $-$0.3 mag$/$dex) are simulated.
Since seeing effects on color gradients may depend on angular size and
ellipticity of galaxy (e.g., Peletier et al. 1990a), the parameters of
the model galaxies are sampled from the ranges covered by the sample
galaxies: apparent magnitude in $R$ band (14, 15, and 16 mag), effective
radius (1$^{\prime\prime}$, 3$^{\prime\prime}$, and 6$^{\prime\prime}$)
and ellipticity (0.1, 0.4, and 0.7). The magnitude and effective radius
were assigned to a model galaxy taking into account the scaling relation
at the distance of the cluster and the typical dispersion around the
relation.
It should be also kept in mind that PSFs are not necessarily circular in
real images. Since such non-circular PSFs may have effects on color
gradients, they are also to be considered in the simulations. In order
to see whether PSFs are circular in our data or not, we picked up bright
stellar objects using SExtractor based on the ``CLASS STAR'' parameter
and examined their shapes. It is found that PSFs tend to be slightly
deviant from circles and the ellipticities are between 0 and 0.1, and a
typical value is $\sim 0.05$. For position angle, it is suggested that
the seeing effects on a color profile are more serious when the position
angle of a non-circular PSF is perpendicular to that of a galaxy (e.g.,
Michard 1999). Accordingly, we assume a PSF with an ellipticity of 0.1
and a position angle perpendicular to that of a model galaxy to simulate
the worse case.

The model galaxies are analyzed in the same way as the real galaxies
including corrections for differences in seeing size between the $B$ and
$R$ images, and their color profiles and color gradients are obtained.
The simulations are executed in all the stacked images. In the
following, several characteristics of the seeing effects on color
profiles and color gradients are mentioned based on the results of the
simulations.

(1) First, basic trends of the seeing effects on radial color profiles
are described in a qualitative way.  If a seeing size in a $B$ band
image is larger than that in an $R$ band image, a color in the innermost
portion after correction for the difference in seeing size between the
images, which is referred to ``convolved color'' hereafter, becomes
redder than the intrinsic color. The convolved color becomes bluer
monotonically with increasing radius, and in the outer region, it
becomes bluer than the intrinsic color. (The turnover occurs around an
effective radius of a galaxy.) In other words, a color gradient steeper
than the intrinsic value would be obtained due to the seeing effects.
If a seeing size in an $R$ band image is larger, the convolved color
becomes slightly bluer than the intrinsic one near the galaxy center
and becomes redder with radius. After it overtakes the intrinsic color,
the color profile has a peak and the convolved color becomes bluer with
radius. Eventually, it becomes bluer than the intrinsic color in the
outer region. Even in this case, a color gradient is estimated to be
steeper than the intrinsic one, unless all the data points including the
innermost one are used to derive a color gradient from the color
profile.

(2) In the following, we mention the seeing effects more quantitatively
together with dependencies of the seeing effects on parameters of the
model galaxies. The deviation of the color gradient affected by seeing
from the intrinsic value is larger for a galaxy with a smaller effective
radius, i.e., with a more compact light distribution. A typical amount
of the deviation is estimated to be $-0.02$ mag/dex, $-0.04$ mag/dex and
$-0.08$ mag/dex for an effective radius of $6^{\prime\prime}$,
$3^{\prime\prime}$, and $1^{\prime\prime}$, respectively.

(3) The deviation is larger for a galaxy with a steeper intrinsic color
gradient. Along with the variation of the intrinsic color gradient from
$-0.1$ to $-0.3$ mag/dex, a typical amount of the deviation increases
from $-0.02$ mag/dex to $-0.08$ mag/dex.

(4) If the intrinsic color gradient and effective radius are fixed to
some values, there is no clear dependency of the seeing effects on
ellipticity of the model galaxy.

Basically, a larger inner cut-off radius tends to make the seeing
effects on a color gradient less serious. However, the number of the
data points with high $S/N$s available to fit the regression line to the
color profile would be also reduced and thus the error in the
measurement of color gradient would be larger. This is likely to be more
serious for the less luminous and thus smaller galaxy. Considering these
effects, we decided to compromise and adopt $1_{\cdot}^{\prime\prime}5$
as the inner cut-off radius. It has to be mentioned that some residual
seeing effects could still exist and the color gradient might be
estimated to be steeper than the intrinsic one. The typical amount of
the deviation is predicted to be $-0.01$ mag$/$dex, $-0.02$ mag$/$dex,
and $-0.06$ mag$/$dex for a galaxy with an effective radius of
$6^{\prime\prime}$, $3^{\prime\prime}$, and $1^{\prime\prime}$,
respectively. It is worth mentioning that each of these deviations is
comparable to or smaller than the typical error in the individual
measurement of color gradient in the model galaxy with each of the
effective radii.

Finally, some comments are mentioned on the three galaxies of which
color gradients were studied by Peletier et al. (1990a) (NGC 5638 and
NGC 5831) and by Franx et al. (1989) (NGC 7619). 
In Figure 3, the radial profiles of $B-R$ color obtained in this study
and those in the previous studies are compared in the upper panels, and
the differences are presented in the lower panels. The uncertainty of
the $\Delta (B-R)$ at each radius is comparable to or larger than the
error bar in the color profile obtained in this study. Color gradients
of these galaxies are re-estimated by using the data for the color
profiles in the literature and the outer cut-off radii defined in this
study, while we adopted the inner cut-off radii defined in the
literature. The resultant color gradients are compared in Table
\ref{comptab}. For NGC 5638 and NGC 5831, the values agree reasonably
with each other. For NGC 7619, although the difference in color gradient
seems to be large, it is unlikely to be significant if the large error
in the color profile presented by Franx et al. (1989) is considered.

\subsection{Statistical Properties of Color Gradients}

In Figure 4, histograms of $B-R$ color gradients of the sample galaxies
in ABELL 2199 and ABELL 2634 are shown. A width of each bin in the
histograms is taken to be 0.05 mag/dex of $d (B-R) / d \log r$, which is
the typical uncertainty of the derived individual color gradient. A red
solid line and a red dashed line depict the histograms for the galaxies
in ABELL 2199 and in ABELL 2634, respectively. We calculated a mean
value of color gradient and a standard deviation of each distribution,
and summarized the results in Table \ref{values}.  The mean value of the
whole sample of ABELL 2199 and ABELL 2634 is $d (B-R) / d \log r =
-0.09$ mag/dex with the standard deviation of $0.03-0.04$. A median
value is also close to this. Although less luminous ($R > 15$ mag) and
smaller ($r_e < 3^{\prime\prime}$) galaxies show steeper color gradients
on average, this is unlikely to be significant in the histogram or the
Table, if the characteristics of the seeing effects found in the
simulations as well as the larger uncertainty in measurement of color
gradient are taken into account.  These results are independent of
whether the ``embedded'' galaxies are excluded from the samples or not.

We next examine the relations between the color gradients and global
properties of the cluster galaxies. In Figure 5, the $B-R$ color
gradients in the sample galaxies are plotted against $R$ band total
magnitudes, effective radii, central velocity dispersions, and Mg$_2$
indices. The $R$ band total magnitude and the effective radius are based
on our data. For the $R$ band total magnitudes, the error includes the
calibration error. For the effective radii, the error is based on 1
$\sigma$ fitting error in slope of the regression line to the surface
brightness profile. The central velocity dispersion, Mg$_2$ index, and
their errors are taken from Lucey et al. (1997).
Open circles denote the galaxies embedded in the envelope of a
neighboring galaxy of which color gradients were obtained after the
neighboring galaxy was modeled and subtracted.
Red symbols indicate the galaxies with both the effective radius larger
than $\sim 3^{\prime\prime}$ and the $R$ band total magnitude brighter
than 15 mag.
As seen in Figure 5, there are no clear correlations when we consider
the whole sample. However, in the relation between the color gradients
and $R$ band total magnitudes of the galaxies in ABELL 2199, it is
suggested that there is a discontinuity in the relation around $R = 15$
mag, that is, the relation in the magnitude range brighter than $R = 15$
mag may be different from that in the fainter magnitude range.  For the
galaxies brighter than $R = 15$ mag, a weak correlation may exist:
brighter galaxies have steeper color gradients. (The correlation
probability is estimated to be larger than 99\% based on the Spearman's
rank correlation coefficient, while the linear correlation coefficient
is estimated to be 0.6.) In the lower luminosity range, on the other
hand, it is hard to see whether the correlation exists or not due to the
larger uncertainties in the color gradients. The $R$ band apparent
magnitude of 15 mag corresponds to an absolute magnitude of $-21.3$ mag
and thus $\sim L^{*}$ (e.g., Lin et al. 1996).

Similar features appear to exist in the relation between the color
gradients and effective radii: there seems to be a break around an
effective radius of $3^{\prime\prime}$, which roughly corresponds to $R
= 15$ mag considering the scaling relation, and for the galaxies with
effective radii larger than $\sim 3^{\prime\prime}$, the galaxies with
larger effective radii tend to have steeper color gradients.
%
%
These trends are preserved if the ``embedded'' galaxies (open circles in
Figure 5) are excluded from the diagrams. Although it is unclear for the
galaxies in ABELL 2634 whether such trends exist or not due to the small
number of the galaxies, the data points appear to be consistent with the
trends seen for the galaxies in ABELL 2199. It should be stressed here
that these trends are not expected by the seeing effects, because the
seeing effects could make color gradients steeper for less luminous and
thus smaller galaxies.
Almost the same trends are seen when the inner cut-off radius is adopted
to be $2^{\prime\prime}$. On the other hand, they disappear when the
cut-off radius is defined to be $\leq 1^{\prime\prime}$, probably
because of the seeing effects.
Although similar trends are expected to be seen in the relations with
the central velocity dispersions and Mg$_2$ indices, no clear trends are
seen. We suggest that this is at least partly due to the relatively
large errors of central velocity dispersion and Mg$_2$ index. In fact,
these quantities from Lucey et al. (1997) only weakly correlate with the
$R$ band total magnitudes and effective radii obtained from our data.

\section{DISCUSSION}

If evolutionary histories of elliptical galaxies depend on galaxy
environment, their metallicity gradients and thus color gradients could
also have some environmental dependences. As described in \S~1, recent
studies either observationally or theoretically suggest that ellipticals
which reside in rich clusters tend to have formed through the formation
processes quite similar to the monolithic collapse, in which more
massive ellipticals have been suggested to have steeper metallicity
gradients (Larson 1974b; Carlberg 1984). On the other hand, elliptical
galaxies in less dense environments are likely to have formed through
major mergers, which are expected to dilute the existing metallicity
gradients and any correlations between metallicity gradients and global
properties of the galaxies.
Consequently, a significant difference in statistical property of color
gradient between elliptical galaxies in rich clusters and those in the
sparse environments is expected.
The motivation of this study is to examine these issues observationally
by adding the data of color gradients in elliptical galaxies in rich
clusters, which have been rarely studied.

\subsection{Comparison with Monolithic Collapse Models}

The mean value of the color gradient obtained in the ellipticals in
ABELL 2199 and ABELL 2634 is $-0.09 \pm 0.04$ mag/dex. Using a simple
stellar population model by Vazdekis et al. (1996), this color gradient
is converted to a metallicity gradient ($d \log Z / d \log r$) of $\sim
-0.3 \pm 0.1$ by assuming an age of 12 Gyr. On the other hand, numerical
simulations based on the monolithic collapse scenario by Carlberg (1984)
and by Kobayashi (2002) predicted a metallicity gradient of $\sim -0.5$
and $\sim -1.0$, respectively, which are significantly steeper than our
estimated metallicity gradient.

However, it has been recently suggested that if a proto-galactic gas
cloud has small scale density fluctuations expected in the CDM universe
rather than a uniform density, an elliptical galaxy could form by the
initial starburst following the successive assembly of low mass and
gas-rich sub-clumps, and the metallicity gradient and $B-R$ color
gradient in the galaxy could be consistent with our results (Kawata
2001). In his simulation, a giant elliptical with a $B$ band absolute
magnitude of $-21.4$ mag at $z = 0$ was studied, and its metallicity
gradient and $B-R$ color gradient were estimated to be $\sim -0.30$ and
$\sim -0.13$ mag/dex, respectively. These values are consistent with the
gradients in the brightest galaxies obtained in our study.
If that process is more realistic in the monolithic collapse, our
results may imply that elliptical galaxies in rich clusters tend to have
formed through a process like the monolithic collapse.

Our data suggest that for the galaxies in the clusters brighter than $R
= 15$ mag and those with effective radii larger than $\sim
3^{\prime\prime}$, more luminous and larger galaxies tend to have
steeper color gradients. This tendency can be seen in the results by
numerical simulation based on the monolithic collapse scenario (Larson
1974b; Carlberg 1984; see also Kawata \& Gibson 2002). Thus it also
seems to support the view that the cluster ellipticals formed by the
monolithic collapse scenario. Since the trend found is rather weak and
may not be very convincing, it is important to study ellipticals in
other rich clusters.

It is interesting that the correlation seen for the galaxies brighter
than $R = 15$ mag does not seem to continue to the lower luminosity
range. It has been suggested that ellipticals less luminous than $L^{*}$
have observational properties (isophotal shape, kinematics, etc)
different from more luminous ones (e.g., Bender, Burstein \& Faber 1992;
Faber et al. 1997) and thus have different formation mechanisms. The
discontinuity found here may be another piece of the evidence, even
though the color gradients in the less luminous galaxies obtained in
this study are still uncertain. Deeper observations in sub-arcsecond
seeing conditions are necessary for more detailed study.

\subsection{Environmental Dependence}

We are now in a position to compare the cluster ellipticals with those
in less dense environments. In Figure 4, the histograms of the color
gradients of the galaxies in ABELL 2199 and ABELL 2634 are compared with
those of 42 nearby elliptical galaxies studied by Michard (1999), 39
elliptical galaxies by Peletier et al. (1990a), and 17 elliptical
galaxies by Franx et al. (1989). Some of the ellipticals studied in the
literature are located in nearby clusters. Most of them are Virgo
ellipticals, and a handful of ellipticals in the other poor clusters
(Fornax, AWM 4, and AWM 5) or rich clusters (Coma cluster, Pegasus
cluster, ABELL 2029, and ABELL 2162) are included. We also calculated
the mean value of the color gradient and the standard deviation in a
sub-sample of the cluster ellipticals, and the results are presented in
Table \ref{values}. The distributions of the color gradients for ABELL
2199 and ABELL 2634 seem to be quite similar to those by Peletier et
al. (1990a) and by Franx et al. (1989), though it is slightly shifted to
a steeper side than the distribution by Michard (1999).
\footnote{It is suggested that for the 19 galaxies studied by both
Michard (1999) and Peletier et al. (1990a), their $B-R$ color gradients
do not well correlate with each other (Michard 1999).}
The similarity in the distribution suggests that distribution of color
gradient is almost independent of galaxy environment.

Next, using the published data, color gradients in nearby elliptical
galaxies are plotted against the $B$ band absolute magnitudes (Figure
6(a)) and effective radii (Figure 6(b)). The $B$ band absolute
magnitudes and effective radii in the diagram by Idiart et al. (2002)
were taken from Michard (1999) and Michard \& Marchal (1994).  In the
other diagrams, the data presented in each literature were used.
Since effective radii measured along major axis are tabulated in
Peletier et al. (1990a) and Goudfrooij et al. (1994), we converted them
to the equivalent radii using the ellipticities in the literature.
Solid circles indicate ellipticals in sparse environments such as field
or group. Open circles and open squares denote ellipticals in Virgo
cluster and those in the other poor clusters, respectively. Open
triangles indicate those in rich clusters. (Most of them are the
brightest cluster galaxies.) It should be emphasized that Virgo cluster
is not a rich cluster but a poor cluster in which a spatial distribution
of galaxies is known to be rather sparse, and color gradients in only a
few ellipticals in a cluster as rich as ABELL 2199 have been studied so
far.
We note that a $B$ band absolute magnitude of $\sim -20$ mag and an
effective radii of 2.5 kpc roughly correspond to an $R$ band magnitude
of 15 mag considering a typical color of an elliptical galaxy and an
angular size of 3$^{\prime\prime}$ at the distance of the clusters.

Figure 6 suggests that, contrary to the galaxies in ABELL 2199 and ABELL
2634, there are no clear trends between the color gradients and galaxy
luminosities or effective radii for ellipticals in the less dense
environments. This is consistent with the conclusion by Kobayashi \&
Arimoto (1999) based on the studies of the radial gradients of metal
absorption line index.
Although there seems to be a marginal trend in the study by Peletier
et al. (1990a) that for the cluster ellipticals (open symbols),
brighter and larger ellipticals have steeper color gradients,
\footnote{This trend would be consistent with those found in ABELL 2199
and ABELL 2634. However, it should be interpreted carefully because the
ellipticals are sampled from several clusters. It seems that the color
gradients of the Virgo ellipticals (open circles) do not strongly
correlate with their luminosities or effective radii. It should be also
noted that the ellipticals in the clusters other than Virgo occupy the
high luminosity end. Therefore, it is not obvious whether such a trend
as that found in ABELL 2199/2634 exists in each cluster.}
it is still worth mentioning that the color gradients in the non-cluster
ellipticals (solid symbols) do not correlate with their global
photometric properties.

In summary, while it is suggested that distribution of color gradient in
dense environment such as in ABELL 2199 is not significantly different
from that in sparse environment, there seems to be an environmental
dependence in relation between color gradients and global properties of
ellipticals: more luminous and larger ellipticals have steeper color
gradients in dense environment, but any such trend is clearly weaker in
less dense environments.
These results do not fully support the view that ellipticals in rich
clusters formed through the monolithic collapse while those in sparse
environments formed through galaxy mergers, because the latter
ellipticals are expected to have color gradients shallower on average
than the formers.
This apparent conflict may be reconciled if star formation and
subsequent chemical enrichment induced by the mergers are taken into
account. It is suggested by numerical simulations that if a galaxy
merger occurs, a significant fraction of gas in the progenitor galaxies
falls into the central regions of the merger remnant and thus star
formation activities are also expected to be more active in the inner
regions (e.g., Barnes \& Hernquist 1991). Consequently, stellar
populations in the inner regions would be more chemically enriched and
thus the merger remnant could acquire a metallicity gradient (Mihos \&
Hernquist 1994). If this effect compensates the dilution of the
metallicity gradient existing in the progenitors by mixing stars during
the merger, a resultant metallicity gradient in the merger remnant would
not become significantly shallower.
Nevertheless, it is still difficult to address which is more important,
the acquisition or the dilution, to determine a metallicity gradient in
a merger remnant, because it is likely to depend on initial condition of
galaxy merger. It is also possible that a galaxy has experienced several
mergers before the present time and thus the metallicity gradient is
defined by its whole merging history.

\subsection{Similarity to Color Gradient in Spiral Bulge}

It is interesting to point out here that a similar trend to that found
in this work has been suggested for a sample of bulges of spiral
galaxies. In Balcells \& Peletier (1994), $U - R$ and $B - R$ color
gradients in 18 nearby spiral bulges were studied. Their sample consists
of bulges of spiral galaxies earlier than Sc, and barred spirals and
dusty spirals are excluded. They found that for the bulges more luminous
than an $R$ band absolute magnitude of $-20$ mag, the bulge luminosities
correlate with the color gradients in the sense that more luminous
bulges have steeper color gradients. No clear correlations were found
between the bulge color gradients and the total luminosities or any of
the parameters related to disk light. Since the $B - R$ color gradients
in the bulges vary from $\sim 0$ mag/dex to $\sim -0.2$ mag/dex for an
increase of the bulge luminosities by 2 magnitudes, the correlation
seems to be quite similar to the trend found in this study both
qualitatively and quantitatively, though the cluster ellipticals studied
here are $\sim 1$ magnitude more luminous on average than the bulges.
It is also suggested by Balcells \& Peletier (1994) that there is a
correlation between the bulge color gradients and bulge luminosities for
the less luminous bulges ($M_{R,Bulge} \geq -20$), but it is opposite to
that for the more luminous ones: less luminous bulges have steeper color
gradients. Although it is unclear in our study whether there is a
correlation for the cluster ellipticals in the low luminosity range, our
data do not exclude the possibility that there is a discontinuity in the
relation around $R = 15$ mag.
These common features may suggest that luminous bulges of spiral
galaxies have formation histories more similar to ellipticals in rich
clusters. However, once the environmental dependence found in this work
for elliptical galaxies is taken into account, it does not seem to be
simply explained because spiral galaxies tend to be in sparse
environments, while no clear trends between color gradients and galaxy
luminosities have been found for ellipticals in sparse
environments. More observational and theoretical works would be
necessary to understand what this similarity implies.

\section{SUMMARY}

We performed $B$ and $R$ band surface photometry for E/S0 galaxies in
ABELL 2199 and ABELL 2634, and studied their $B-R$ color gradients. In
order to sample spheroid dominated galaxies, the decomposition of the
surface brightness profiles into $r^{1/4}$ bulge and exponential disk
profiles was performed and the sample selection was based on the $B$
band bulge-to-total luminosity ratios. Resultantly, 29 galaxies in ABELL
2199 and 7 galaxies in ABELL 2634 were sampled. We evaluated the seeing
effects on the color gradient by simulating with artificial galaxies
having various parameters and defined the inner cut-off radius in
deriving color gradients to mitigate the effects.

The mean value of the color gradient is estimated to be $-0.09 \pm 0.04$
mag/dex, which can be converted to a metallicity gradient of $\sim -0.3
\pm 0.1$. This gradient seems to be consistent with that expected by a
recent monolithic collapse model. We further studied the relations
between the $B-R$ color gradients and global properties of the sample
galaxies. It is found that for the galaxies in ABELL 2199 brighter than
$R = 15$ mag, which roughly corresponds to $L^{*}$ at the distance of
the cluster, there may be a correlation in the sense that more luminous
galaxies have steeper color gradients. Also, for the galaxies with
effective radii larger than $\sim 3^{\prime\prime}$, which nearly
corresponds to $R = 15$ mag considering the scaling relation, the
galaxies with larger effective radii seem to have steeper color
gradients. The ellipticals in ABELL 2634 appear to follow the same
trends, though the sample is small.  These trends also suggest that
elliptical galaxies in rich clusters formed through the monolithic
collapse scenario. Since this is the first systematic study of color
gradients in ellipticals in a rich cluster, it is important to study
ellipticals in other rich clusters and to see whether the
characteristics of the color gradients found in this study are generally
seen or not.

We compared the results with the color gradients in elliptical galaxies
in less dense environments using data from the literature. It is found
that no clear trends between color gradient and galaxy luminosity or
effective radius are seen, suggesting that elliptical galaxies in sparse
environments formed through galaxy mergers, while the distributions of
the color gradients seem to be quite similar to those obtained in ABELL
2199 and ABELL 2634.
These results suggest that while distribution of color gradient is
almost independent of galaxy environment, there may be an environmental
dependence in relation between color gradients and global properties of
ellipticals: more luminous and larger ellipticals have steeper color
gradients in dense environment, but any such trend is clearly weaker in
less dense environments.
This does not fully support the view that ellipticals in rich clusters
formed through the monolithic collapse while those in sparse
environments formed through galaxy mergers, because the latter
ellipticals are expected to have color gradients shallower on average
than the formers.
This apparent conflict may be reconciled if we take into account star
formation and subsequent chemical enrichment induced by the mergers.
Since the chemical enrichment could proceed more in the inner regions of
the merger remnant, a metallicity gradient could be newly formed and it
may compensate the dilution of the metallicity gradient.

The correlation between the color gradient and the galaxy luminosity (or
the effective radius) found for the brighter cluster ellipticals cannot
be seen for the fainter ellipticals; this may imply the difference of
formation process between them, though it should be confirmed by data
for the fainter ellipticals with a higher spatial resolution.
On the other hand, the relation seems to be similar, both qualitatively
and quantitatively, to that previously found for luminous bulges of
spiral galaxies. This may indicate that cluster ellipticals and bulges
of spiral galaxies have similar formation mechanisms.

\acknowledgements

We appreciate the support from members of the University of Hawaii
observatory during the observations.  We are grateful to the anonymous
referee for useful comments and suggestions to improve our paper. We
would like to thank C. Kobayashi and N. Arimoto for fruitful
discussions. This research made use of the NASA/IPAC Extragalactic
Database (NED), which is operated by the Jet Propulsion Laboratory,
California Institute of Technology, under a contract with the National
Aeronautics and Space Administration.

\clearpage

\centerline{\bf Figure Caption}

\noindent
Figure 1 --- A top panel shows $R$ band image of a sample galaxy. In
Figure 1(a) and 1(b), the sample galaxies in ABELL 2199 and those in
ABELL 2634 are presented in order of apparent magnitude in $R$ band,
respectively. The three calibration galaxies (NGC 5638, NGC 5831, and
NGC 7619) are presented in Figure 1(c). Value of $B/T$ and morphology
are indicated in parentheses.  In a middle panel, azimuthally averaged
surface brightness profiles in $B$ band (open circles) and in $R$ band
(filled circles) are shown. The data points in the $R$ band profile
between the two vertical lines are used to fit a regression line to
derive an effective radius of a galaxy. The best-fit line and the lines
with the slopes $\pm 1\sigma$ deviant from the best-fit are
shown. Dashed line indicates the outer cutoff radius in deriving a color
gradient. In a bottom panel, a radial profile of $B-R$ color is
shown. Abscissa refers to a logarithmic radius along major axis. The two
vertical lines indicate the inner and outer cutoff radii. The regression
line best fit to the color profile and the lines with the slopes $\pm
1\sigma$ deviant from the best-fit are shown.

\noindent
Figure 2 --- Color maps of the sample galaxies.  In Figure 2(a), the
sample galaxies in ABELL 2199 are presented, and in Figure 2(b), those
in ABELL 2634 are shown. The three galaxies (NGC 5638, NGC 5831, and NGC
7619) are presented in Figure 2(c). These maps are put in order of $R$
band magnitude of a galaxy from left to right and from top to bottom.  A
cross mark in each panel shows the centroid of the $R$ band image. These
are the {\it pseudo} color maps, and a range of color adopted in each
map, which is shown by the wedge on the right side of the panel, is
different from galaxy to galaxy. We note that vertical lines seen in BO
24 and NGC 7720 are artificial and are caused by bad columns on the CCD.

\noindent
Figure 3 --- Comparison of radial color profile for NGC 5638, NGC 5831,
and NGC 7619. Two panels are shown for each galaxy. In the upper panels,
the $B-R$ color profiles obtained in this study are shown by filled
circles with error bars, and those taken from the previous studies
(Peletier et al. (1990a) for NGC 5638 and NGC 5831 and Franx et al.
(1989) for NGC 7619) are shown by open circles. Zero points of the color
profiles in the previous studies are arbitrarily shifted for clearer
presentation. In the lower panel, a color difference between the two
data at each radius is shown.
The uncertainty of the $\Delta (B-R)$ can be comparable to or larger
than the error bars attached to the color profile obtained in this
study.
The two vertical lines indicate the inner and outer cutoff radii used to
derive a color gradient. One of them at a smaller radius shows the inner
cut-off radius defined in the literature. Another one indicates the
outer cutoff radius defined in this paper. Note that a mean value of
the difference, which is presumably caused by the zero-point error of
our photometry, was removed because it has no effect on color gradient.

\noindent
Figure 4 --- Histograms of color gradients. Red solid line and red
dashed line indicate histograms for the sample galaxies in ABELL 2199
and in ABELL 2634, respectively. Blue line, green line, and cyan line
show histograms for the galaxies studied by Michard (1999), by Peletier
et al. (1990a) and by Franx et al. (1989), respectively. A width of each
bin in the histograms is taken to be 0.05 mag/dex of $d (B-R) / d \log
r$, which is comparable to an error typically seen in an individual
measurement of $B-R$ color gradient in our data.

\noindent
Figure 5 --- Relations between color gradients and global properties of
the sample galaxies. Left panels are the diagrams for the galaxies in
ABELL 2199, and right panels are for the galaxies in ABELL 2634. The
total magnitudes and effective radii are based on our data. The central
velocity dispersions and Mg$_2$ indices are taken from Lucey et al.
(1997). Open stars denote the brightest cluster galaxies (NGC 6166 in
ABELL 2199 and NGC 7720 in ABELL 2634). Open circles indicate the
galaxies embedded in the envelope of a neighboring galaxy and their
color gradients were obtained after the neighbor was modeled and
subtracted.
Red symbols show the galaxies with both an effective radius larger than
$\sim 3^{\prime\prime}$ and an $R$ band total magnitude brighter than 15
mag.
Solid lines in the diagrams of the color gradients and total magnitudes
and effective radii for the galaxies in ABELL 2199 are shown as
references of the trends. In the relation between color gradient and
effective radius, the brightest cluster galaxy is not taken into account
in the reference line because it seems to be deviant from the trend. The
same lines are also shown in the right panels for comparison.

\noindent
Figure 6 --- Relation between color gradients and $B$ band absolute
magnitudes (Figure 6(a)) and that between color gradients and effective
radii (Figure 6(b)) for nearby elliptical galaxies based on the
published data. The data are basically taken from each literature, but
the $B$ band absolute magnitudes and effective radii in the diagram by
Idiart et al. (2002) are taken from Michard (1999) and Michard \&
Marchal (1994). Solid circles indicate ellipticals in sparse
environments such as field or group. Open circles indicate ellipticals
in Virgo cluster and open squares are ellipticals in the other poor
clusters (Fornax, AWM 4, and AWM 5). Open triangles indicate those in
rich clusters (Coma cluster, Pegasus cluster, ABELL 2029, and ABELL
2162).
It is noted that one Virgo elliptical and one Fornax elliptical in
Goudfrooij et al. (1994) are not included here because their color
gradients are studied in $B-V$ or $V-I$ color, not in $B-I$ color
as the other ellipticals are studied.

\clearpage

\begin{deluxetable}{ccccccl}
\tablewidth{0pt}
\tablecaption{Observing log. \label{list1}}
\tablehead{
\multicolumn{2}{c}{Field Center}                      &
\multicolumn{2}{c}{Total Exp. Time}   & \multicolumn{2}{l}{Seeing size} &
\multicolumn{1}{c}{Objects} \\
\colhead{$\alpha$(J2000)} & \colhead{$\delta$(J2000)} &
\colhead{$B$}     & \colhead{$R$}   & \colhead{$B$} & \colhead{$R$} &
\colhead{} \\
\colhead{(h m s)} & \colhead{($^{\circ}~^{\prime}~^{\prime\prime}$)} &
\colhead{(sec)}   & \colhead{(sec)}   & \colhead{($^{\prime\prime}$)} & \colhead{($^{\prime\prime}$)} &
\colhead{} \\
\colhead{(1)}             & \colhead{(2)}             &
\colhead{(3)}     & \colhead{(4)}     & \colhead{(5)} & \colhead{(6)} &
\multicolumn{1}{c}{(7)}
}
\startdata

\multicolumn{7}{c}{ABELL 2199} \\ \tableline
16 28 36.7 & 39 33 58 & 1260 & 750 & 1.1 & 0.9 & NGC 6166 (9100), BO 15 (8782),\\
           &          &      &     &     &     & BO 24 (10182), BO 41 (9323),\\
           &          &      &     &     &     & BO 66 (9209), BO 73 (7993),\\
           &          &      &     &     &     & BO 74 (8548), BO 95 (8259)\\ \tableline
16 27 46.7 & 39 34 21 & 1860 & 780 & 1.1 & 0.9 & BO 25 (9372), BO 46 (9149) \\ \tableline
16 29 33.9 & 39 48 52 & 1860 & 780 & 1.0 & 1.2 & BO 19 (7833), BO 26 (9176) \\ \tableline
16 28 53.4 & 39 28 01 & 1920 & 750 & 0.9 & 0.9 & BO 34 (8679), BO 51 (10164) \\  \tableline
16 28 59.6 & 39 41 26 & 1800 & 720 & 1.2 & 1.1 & BO 38 (8397), BO 45 (9400),\\
           &          &      &     &     &     & BO 107 (9163), BO 113 (8477) \\ \tableline
16 27 40.0 & 39 23 00 & 1260 & 600 & 1.1 & 1.0 & NGC 6158 (8980), BO 70 (8777),\\
           &          &      &     &     &     & BO 96 (9253) \\ \tableline
16 27 52.9 & 39 15 33 & 1800 & 600 & 1.0 & 1.0 & BO 5 (8744), BO 20 (9638),\\
           &          &      &     &     &     & BO 69 (8956) \\ \tableline
16 27 03.7 & 39 31 57 & 1800 & 960 & 1.1 & 0.8 & BO 8 (10159) \\ \tableline
16 25 40.0 & 39 36 20 &  660 & 300 & 1.1 & 1.2 & RS 8 (8952) \\ \tableline
16 28 18.3 & 38 48 43 & 1860 & 600 & 1.1 & 1.1 & RS 72 (9288) \\ \tableline
16 31 18.5 & 39 08 53 & 1500 & 600 & 0.9 & 1.0 & RS 162 (8856), RS 163 (8854) \\ \tableline
           &          &      &     &     &     & \\
\multicolumn{7}{c}{ABELL 2634} \\ \tableline		      
23 38 28.8 & 27 01 53 &  540 & 180 & 0.9 & 0.9 & NGC 7720 (9060), IC 5342 (9274),\\
           &          &      &     &     &     & BO 11 (9502), BO 27 (8378),\\
           &          &      &     &     &     & BO 82 (9552), D 75 (9828),\\
           &          &      &     &     &     & LGC 125 (9983) \\ \tableline
           &          &      &     &     &     & \\
\multicolumn{7}{c}{Others} \\ \tableline
14 29 40.4 & 03 14 00 &  360 & 150 & 0.8 & 0.9 & NGC 5638 (1676) \\ \tableline
15 04 07.0 & 01 13 11 &  420 & 120 & 0.9 & 0.8 & NGC 5831 (1656) \\ \tableline
23 20 14.7 & 08 12 22 &  540 & 150 & 0.9 & 0.8 & NGC 7619 (3762) \\ \tableline
\enddata

\tablecomments{Col. (5) and (6): Seeing size estimated in a stacked
 image. Col. (7): Sample galaxies in each observed field are shown.
 Galaxy IDs with ``BO'', ``D'', ``LGC'', and ``RS'' are based on those
 listed in Butcher \& Oemler (1985), Dressler (1980), Lucey et al.
 (1991), and Rood \& Sastry (1972), respectively. Number in parentheses
 indicates galaxy redshift ($cz$) in km/s.}

\end{deluxetable}

\setlength{\tabcolsep}{1mm}
\begin{deluxetable}{lccccccccc}
\tablewidth{-5pt}
\tablecaption{Properties of the sample galaxies. \label{list2}}
\tablehead{
\colhead{Galaxy} & \colhead{$B/T$} & \colhead{Mor.} &
\colhead{$B$}   & \colhead{$R$}   & \colhead{$(B-R)_e$}  &
\colhead{$\log r_{e}$}  & \colhead{Mg$_{2}$} &
\colhead{$\log \sigma_0$} & \colhead{$d(B-R)/d\log r$}
\\
\colhead{}       & \colhead{}      & \colhead{}     &
\colhead{(mag)} & \colhead{(mag)} & \colhead{(mag)}       &
\colhead{(arcsec)}      & \colhead{(mag)}     &
\colhead{(km/s)}          & \colhead{(mag/dex)}
\\
\colhead{(1)}    & \colhead{(2)}   & \colhead{(3)}  &
\colhead{(4)}    & \colhead{(5)}  & \colhead{(6)}         &
\colhead{(7)}           & \colhead{(8)}       &
\colhead{(9)}             & \colhead{(10)}
}
\startdata 
NGC 6166 & 1.0 & cD     & 14.9 & 13.2 & 1.59 & 1.66 & 0.319    & 2.475    & $-0.12 \pm 0.01$ \\
BO 15    & 0.9 & E      & 15.9 & 14.3 & 1.59 & 0.88 & 0.287    & 2.240    & $-0.09 \pm 0.01$ \\ 	     
BO 24    & 0.7 & S0     & 16.1 & 14.3 & 1.69 & 0.55 & 0.325    & 2.459    & $-0.06 \pm 0.03$ \\ 	     
BO 41    & 0.8 & E/S0   & 16.7 & 15.0 & 1.67 & 0.14 & $\cdots$ & $\cdots$ & $-0.07 \pm 0.04$ \\   
BO 66    & 0.6 & S0     & 16.8 & 15.2 & 1.68 & 0.05 & 0.300    & 2.282    & $-0.19 \pm 0.03$ \\	     
BO 73    & 0.8 & E      & 17.1 & 15.5 & 1.61 & 0.18 & 0.269    & 2.220    & $-0.14 \pm 0.05$ \\ 	     
BO 74    & 0.8 & E      & 16.4 & 14.9 & 1.63 & 0.14 & 0.267    & 2.391    & $-0.07 \pm 0.03$ \\ 	     
BO 95    & 0.8 & E      & 17.0 & 15.4 & 1.61 & 0.32 & 0.275    & 2.195    & $-0.06 \pm 0.06$ \\ \tableline
BO 25    & 0.9 & E      & 16.2 & 14.6 & 1.55 & 0.73 & $\cdots$ & $\cdots$ & $-0.04 \pm 0.02$ \\   
BO 46    & 0.6 & E      & 16.6 & 14.9 & 1.70 & 0.26 & $\cdots$ & $\cdots$ & $-0.04 \pm 0.04$ \\ \tableline
BO 19    & 0.8 & S0     & 16.1 & 14.3 & 1.77 & 0.40 & 0.266    & 2.380    & $-0.14 \pm 0.02$ \\
BO 26    & 0.9 & E/S0   & 16.2 & 14.6 & 1.71 & 0.48 & 0.275    & 2.310    & $-0.13 \pm 0.03$ \\ \tableline
BO 34    & 1.0 & E/S0   & 16.4 & 15.0 & 1.53 & 0.80 & 0.222    & 1.995    & $-0.08 \pm 0.03$ \\
BO 51    & 0.8 & S0     & 17.1 & 15.6 & 1.54 & 0.39 & $\cdots$ & $\cdots$ & $-0.15 \pm 0.05$ \\ \tableline
BO 38    & 0.8 & E      & 17.0 & 14.9 & 2.15 & 0.49 & 0.300    & 2.243    & $-0.03 \pm 0.02$ \\
BO 45    & 0.7 & E      & 17.0 & 14.9 & 2.19 & 0.49 & 0.300    & 2.252    & $-0.05 \pm 0.03$ \\
BO 107   & 0.6 & S0$^*$ & 18.0 & 15.9 & 2.19 & $-$0.08 & $\cdots$ & $\cdots$ & $-0.14 \pm 0.08$ \\
BO 113   & 0.8 & E$^*$  & 18.2 & 16.2 & 2.10 & 0.53 & $\cdots$ & $\cdots$ & $-0.10 \pm 0.09$ \\ \tableline
NGC 6158 & 1.0 & E      & 15.3 & 13.6 & 1.70 & 1.02 & 0.277    & 2.280    & $-0.12 \pm 0.01$ \\
BO 70    & 0.8 & S0$^*$ & 17.2 & 15.6 & 1.58 & 0.30 & $\cdots$ & $\cdots$ & $-0.08 \pm 0.06$ \\
BO 96    & 0.7 & E$^*$  & 17.5 & 15.9 & 1.70 & 0.14 & $\cdots$ & $\cdots$ & $-0.13 \pm 0.08$ \\ \tableline
BO 5     & 0.9 & E      & 15.6 & 13.9 & 1.62 & 0.96 & 0.284    & 2.312    & $-0.14 \pm 0.02$ \\
BO 20    & 0.9 & E      & 16.1 & 14.4 & 1.67 & 0.59 & 0.304    & 2.274    & $-0.06 \pm 0.03$ \\
BO 69    & 0.7 & S0     & 16.3 & 14.6 & 1.70 & 0.39 & $\cdots$ & $\cdots$ & $-0.04 \pm 0.03$ \\ \tableline
BO 8     & 0.9 & E      & 15.5 & 14.0 & 1.55 & 0.99 & 0.269    & 2.146    & $-0.08 \pm 0.01$ \\ \tableline
RS 8     & 0.9 & E      & 15.4 & 13.8 & 1.56 & 0.76 & 0.297    & 2.395    & $-0.09 \pm 0.02$ \\ \tableline
RS 72    & 0.9 & E      & 15.6 & 13.9 & 1.72 & 0.74 & 0.304    & 2.285    & $-0.10 \pm 0.02$ \\ \tableline
RS 162   & 1.0 & E      & 15.8 & 14.1 & 1.66 & 0.77 & 0.288    & 2.378    & $-0.03 \pm 0.01$ \\
RS 163   & 0.6 & S0     & 16.1 & 14.4 & 1.71 & 0.33 & $\cdots$ & $\cdots$ & $-0.11 \pm 0.02$ \\ \tableline
NGC 7720 & 0.9 & E      & 15.0 & 13.3 & 1.76 & 1.11 & 0.332    & 2.520    & $-0.10 \pm 0.01$ \\
IC 5342  & 0.9 & E      & 15.9 & 14.2 & 1.77 & 0.81 & 0.301    & 2.345    & $-0.07 \pm 0.03$ \\
BO 11    & 0.9 & E      & 16.0 & 14.2 & 1.79 & 0.95 & 0.324    & 2.335    & $-0.09 \pm 0.03$ \\
BO 27    & 0.9 & S0     & 16.3 & 14.5 & 1.81 & 0.72 & 0.302    & 2.316    & $-0.04 \pm 0.04$ \\
BO 82    & 0.9 & E      & 17.1 & 15.4 & 1.80 & 0.34 & 0.284    & 2.263    & $-0.12 \pm 0.10$ \\
D 75     & 0.7 & E      & 17.1 & 15.3 & 1.79 & 0.39 & 0.297    & 2.273    & $-0.12 \pm 0.07$ \\
LGC 125  & 1.0 & E$^*$  & 17.9 & 16.2 & 1.67 & 1.08 & $\cdots$ & $\cdots$ & $-0.07 \pm 0.23$ \\ \tableline
NGC 5638 & 0.9 & E      & 13.5 & 11.9 & 1.51 & 1.48 & 0.317    & 2.201    & $-0.09 \pm 0.01$ \\ \tableline
NGC 5831 & 0.8 & E      & 13.6 & 12.0 & 1.62 & 1.44 & 0.289    & 2.220    & $-0.12 \pm 0.01$ \\ \tableline
NGC 7619 & 1.0 & E      & 13.5 & 11.8 & 1.67 & 1.30 & 0.336    & 2.528    & $-0.08 \pm 0.01$ \\
\enddata

\tablecomments{Col. (2): Bulge-to-total luminosity ratio obtained by the
 decomposition of a $B$ band surface brightness profile of a galaxy.
 Col. (3): Morphological types tabulated by Lucey et al. (1997).
 Asterisks are attached to the galaxies of which morphological types are
 assigned by us. Col. (6): Color within an aperture of which radius is
 an effective radius of a galaxy. Col. (8) and (9): Values for the
 cluster galaxies are taken from Lucey et al.  (1997), and those for NGC
 5638, NGC 5831, and NGC 7619 are from Kobayashi \& Arimoto
 (1999). Col. (10): A $B-R$ color gradient in each object is listed with
 $1\sigma$ of the fitting error in deriving the color gradient from a
 radial color profile.}

\end{deluxetable}

\clearpage

\begin{deluxetable}{lccl}
\tablewidth{0pt}
\tablecaption{Comparison of $d (B-R)/ d \log r$ with the previous
 studies. \label{comptab}}
\tablehead{
\colhead{Galaxy} & \colhead{This study} & \colhead{Previous studies} &
\multicolumn{1}{c}{Ref.} \\
\colhead{}       & \colhead{(mag/dex)}  & \colhead{(mag/dex)}      &
\colhead{} \\
\colhead{(1)}    & \colhead{(2)}        & \colhead{(3)}            &
\multicolumn{1}{c}{(4)}
}
\startdata 
 NGC 5638 & $-0.09 \pm 0.01$ & $-0.08 \pm 0.02$ & Peletier et al. (1990a) \\
 NGC 5831 & $-0.12 \pm 0.01$ & $-0.08 \pm 0.02$ & Peletier et al. (1990a) \\
 NGC 7619 & $-0.08 \pm 0.01$ & $-0.03 \pm 0.29$ & Franx et al. (1989) \\
\enddata

\end{deluxetable}

\setlength{\tabcolsep}{2mm}
\begin{deluxetable}{llccc}
\tablewidth{0pt}
\tablecaption{List of average color gradient. \label{values}}
\tablehead{
\multicolumn{2}{c}{Sample} & \colhead{Number} & 
\colhead{$<d (B-R) / d \log r>$} & \colhead{Error}
\\
\colhead{}  & \colhead{}   & \colhead{}       &
\colhead{(mag/dex)} & \colhead{(mag/dex)}
\\
\multicolumn{2}{c}{(1)}    & \colhead{(2)}    & 
\colhead{(3)} & \colhead{(4)}
}
\startdata 
 ABELL 2199 & (all)             & 29 & $-0.09 \pm 0.04$ & 0.03 \\
            & ($R < 15$ mag) & 20 & $-0.08 \pm 0.04$ & 0.02 \\
            & ($R > 15$ mag) &  9 & $-0.12 \pm 0.04$ & 0.06 \\
            & ($r_e > 3^{\prime\prime}$) & 16 & $-0.08 \pm 0.04$ & 0.03 \\
            & ($r_e < 3^{\prime\prime}$) & 13 & $-0.11 \pm 0.05$ & 0.05 \\
 ABELL 2634 & (all)             &  7 & $-0.09 \pm 0.03$ & 0.07 \\ \hline
 Michard et al. (1999)  & (all)      & 42 & $-0.07\pm0.04$ & $-$ \\
                        & (Virgo)    & 15 & $-0.06\pm0.03$ & $-$ \\
 Peletier et al. (1990) & (all)      & 39 & $-0.09\pm0.07$ & 0.03 \\
                        & (Virgo)    & 10 & $-0.05\pm0.02$ & 0.02 \\
                        & (RC)       &  5 & $-0.09\pm0.05$ & 0.05 \\
 Franx et al. (1989)    & (all)      & 17 & $-0.07\pm0.04$ & 0.05 \\
                        & (PC)       &  5 & $-0.06\pm0.05$ & 0.06 \\
\enddata

\tablecomments{Col. (1): ``RC'' indicates a subsample consisting of
 several ellipticals in rich clusters. Several ellipticals not in Virgo
 cluster but in other poor clusters were studied by Franx et al. (1989),
 and ``PC'' indicates a subsample of the galaxies.  Col. (2): The number
 of galaxies in each sample. Col. (3): Mean value of color gradient and
 standard deviation in distribution of color gradient are
 indicated. Col. (4): An error typically seen in an individual
 measurement of color gradient is indicated.}

\end{deluxetable}


\clearpage

\begin{thebibliography}{}

\bibitem[]{}
Balcells, M., \& Peletier, R. F. 1994, AJ, 107, 135

\bibitem[]{}
Barnes, J. E., \& Hernquist L. 1991, ApJL, 370, L65

\bibitem[]{} 
Baugh, C. M., Cole, S., \& Frenk, C. S. 1996, MNRAS, 283, 1361

\bibitem[]{}
Bekki, K., \& Shioya, Y. 1999, ApJ, 513, 108

\bibitem[]{}
Bender, R., Burstein, D., \& Faber, S. M. 1992, ApJ, 399, 462

\bibitem[]{}
Bertin, E., \& Arnouts, S. 1996, A\&AS, 117, 393

\bibitem[]{}
Bower, R. G., Lucey, J. R., \& Ellis, R. S. 1992, MNRAS, 254, 601

\bibitem[]{}
Butcher, H. R., \& Oemler, A., Jr. 1985, ApJS, 57, 665

\bibitem[]{}
Buote, D. A., \& Canizares, C. R. 1996, ApJ, 457, 565

\bibitem[]{}
Burns, J. O., Roettiger, K., Ledlow, M., \& Klypin, A. 1994, ApJL, 427,
L87

\bibitem[]{}
Caldwell, N., Rose, J. A., Sharples, R. M., Ellis, R. S., \& Bower,
R. G. 1993, AJ, 106, 473

\bibitem[]{}
Carlberg, R. G. 1984, ApJ, 286, 403

\bibitem[]{}
Carollo, C. M., Danziger, I. J., \& Buson, L. 1993, MNRAS, 265, 553

\bibitem[]{}
Davies, R. L., Sadler, E. M., \& Peletier, R. F. 1993, MNRAS, 262, 650

\bibitem[]{}
Dressler, A. 1980, ApJS, 42, 565

\bibitem[]{}
Faber, S. M., Tremaine, S., Ajhar, E. A., Byun, Y. -I., Dressler, A.,
Gebhardt, K. Grillmair, K., Kormendy, J., Lauer, T. R., \& Richstone, D.
1997, AJ, 114, 1771

\bibitem[]{}
Franx, M., Illingworth, G., \& Heckman, T. 1989, AJ, 98, 538

\bibitem[]{}
Gonz\'{a}lez, J. J. 1993, PhD Thesis, Univ. of California

\bibitem[]{}
Goudfrooij, P. \& de Jong, T. 1995, A\&A, 298, 784

\bibitem[]{}
Goudfrooij, P., Hansen, L., J{\o}rgensen, H. E., N{\o}rgaard-Nielsen, H. 
U., de Jong, T., \& van den Hoek, L. B.  1994, A\&AS, 104, 179

\bibitem[]{}
Governato, F., Gardner, J. P., Stadel, J., Quinn, T., \& Lake, G. 1999,
ApJ, 117, 1651

\bibitem[]{}
Hinkley, S., \& Im, M. 2001, ApJ, 560, L41

\bibitem[]{}
Idiart, T. P., Michard, R., \& de Freitas Pacheco, J. A., 2002, A\&A,
383, 301

\bibitem[]{}
Kauffmann, G. 1996, MNRAS, 281, 487

\bibitem[]{}
Kauffmann, G., \& Charlot, S. 1998, MNRAS, 294, 705

\bibitem[]{}
Kawata, D. 1999, PASJ, 51, 931

\bibitem[]{}
Kawata, D. 2001, ApJ, 558, 598

\bibitem[]{}
Kawata, D., \& Gibson, B. K. 2002, MNRAS, accepted (astro-ph/0212401)

\bibitem[]{}
Kobayashi, C., \& Arimoto, N. 1999, ApJ, 527, 573

\bibitem[]{} 
Kobayashi, C. 2002, PhD Thesis, University of Tokyo

\bibitem[]{}
Kodama, T. 1997, PhD Thesis, Univ. of Tokyo

\bibitem[]{}
Kodama, T., \& Arimoto, N. 1997, A\&A, 320, 41

\bibitem[]{}
Kodama, T., Arimoto, N., Barger, A. J., \& Arag\'{o}n-Salamanca,
A. 1998, A\&A, 334, 99

\bibitem[]{}
Landolt, A. U. 1992, AJ, 104, 340

\bibitem[]{}
Larson, R. B. 1974a, MNRAS, 166, 585

\bibitem[]{}
Larson, R. B. 1974b, MNRAS, 169, 229

\bibitem[]{}
Lin, H., Kirshner, R. P., Shectman, S. A., Landy, S. D., Oemler, A.,
Tucker, D. L., \& Schechter, P. L. 1996, ApJ, 464, 60

\bibitem[]{}
Lucey, J. R., Gray, P. M., Carter, D., \& Terlevich, R. J. 1991, MNRAS, 
248, 804

\bibitem[]{}
Lucey, J. R., Guzm\'{a}n, R., Steel, J., \& Carter, D. 1997, MNRAS, 287, 
899

\bibitem[]{}
Martinelli, A., Matteucci, F., \& Colafrancesco, S. 1998, MNRAS, 298, 42

\bibitem[]{}
Mehlert, D., Saglia, R. P., Bender, R., \& Wegner, G. 2000, A\&AS 141, 449

\bibitem[]{}
Menanteau, F., Abraham, R. G., \& Ellis, R. S. 2001, MNRAS, 322, 1

\bibitem[]{}
Menanteau, F., Ellis, R. S., Abraham, R. G., Barger, A. J., \& Cowie, L. 
L. 1999, MNRAS, 309, 208

\bibitem[]{}
Michard, R. 1999, A\&AS, 137, 245

\bibitem[]{}
Michard, R., \& Marchal, J. 1994, A\&AS, 105, 481

\bibitem[]{}
Mihos, J. C., \& Hernquist, L. 1994, ApJ, 427, 112

\bibitem[]{}
Peacock, J. A. 1999, Cosmological Physics (Cambridge: Cambridge
University Press)

\bibitem[]{}
Peletier, R. F., Davies, R. L., Illingworth, G. D., Davis, L. E., \&
Cawson ,M. 1990a, AJ, 100, 1091

\bibitem[]{}
Peletier, R. F., Valentijn, E. A., \& Jameson, R. F. 1990b, A\&A, 233, 62 

\bibitem[]{}
Pinkney, J., Rhee, G., Burns, J. O., Hill, J. M., Oegerle, W., Batuski,
D., \& Hintzen, P. 1993, 416, 36

\bibitem[]{}
Pritchet, C., \& Kline, M. I. 1981, AJ, 86, 1859

\bibitem[]{}
Rood, H. J., \& Sastry, G. N. 1972, AJ, 77, 451

\bibitem[]{}
Saglia, R. P., Maraston, C., Greggio, L., Bender, R., \& Ziegler,
B. 2000, A\&A, 360, 911

\bibitem[]{}
Shioya, Y., \& Bekki, K. 1998, ApJ, 504, 42

\bibitem[]{}
Simien, F., \& de Vaucouleurs, G. 1986, ApJ, 302, 564

\bibitem[]{}
Stanford, S. A., Eisenhardt, P. R., \& Dickinson, M. 1998, ApJ, 492, 461

\bibitem[]{}
Tantalo, R., Chiosi, C., Bressan, A., Marigo, P., \& Portinari, L. 1998,
A\&A, 335, 823 

\bibitem[]{}
Tamura, N., Kobayashi, C., Arimoto, N., Kodama, T., \& Ohta, K. 2000,
AJ, 119, 2134

\bibitem[]{}
Tamura, N., \& Ohta, K. 2000, AJ, 120, 533

\bibitem[]{}
Terlevich, A. I., \& Forbes, D. A. 2002, MNRAS, 330, 547

\bibitem[]{}
Vader, J. P., Vigroux, L., Lachi\`{e}ze-Rey, M., \& Souviron, J. 1988,
A\&A , 203, 217

\bibitem[]{}
Vazdekis, A., Casuso, E., Peletier, R. F., \& Beckman, J. E. 1996, ApJ,
106, 307

\bibitem[]{}
White, S. D. M. 1980, MNRAS, 191, 1

\bibitem[]{}
Wise, M., \& Silva, D. R. 1996, ApJ, 461, 155

\bibitem[]{}
Witt, A. N., Thronson, H. A. Jr., \& Capuano, J. M. 1992, ApJ, 393, 611

\end{thebibliography}
\end{document}